\documentclass[prb,reprint,aps,floatfix,showpacs,superscriptaddress,longbibliography]{revtex4-1}
\usepackage{amssymb}
\usepackage{bbm}
\usepackage{amsmath}
\usepackage{graphicx}
\usepackage[caption=false]{subfig}
\usepackage[colorlinks=true,linkcolor=blue,anchorcolor=red,citecolor=blue,urlcolor=blue]{hyperref}

\begin{document}

\title{Simulating Cosmological Evolution by Quantum Quench of an Atomic BEC}

\date{\today}
\author{Ke Wang}
\affiliation{Department of Physics and James Franck Institute, University of Chicago, Chicago, Illinois 60637, USA}
\affiliation{Kadanoff Center for Theoretical Physics, University of Chicago, Chicago, Illinois 60637, USA}
\author{Han Fu}
\affiliation{Department of Physics, College of William and Mary, Williamsburg, Virginia 23187, USA
}
\author{K. Levin}
\affiliation{Department of Physics and James Franck Institute, University of Chicago, Chicago, Illinois 60637, USA}
\begin{abstract}
In cosmological evolution, it is the
homogeneous scalar field (inflaton) that drives the universe to expand isotropically and to generate standard model particles. However, to simulate cosmology, atomic gas research has focused on the dynamics of Bose-Einstein condensates
(BEC) with continuously applied forces.
In this paper we argue a complementary approach
needs also to be pursued; we, thus, consider
the analogue BEC experiments in a non-driven, closed
atomic system. We implement this using a BEC in an optical lattice
which, after a quench, freely transitions from an unstable to a stable
state.
This dynamical evolution
displays the counterpart
``preheating", ``reheating" and
``thermalization" phases of cosmology.
Importantly, our studies of these analogue processes yield tractable
analytic models. Of great utility to the cold atom community,
such understanding elucidates the dynamics of non-adiabatic condensate
preparation.
\end{abstract}

\maketitle

\textit{Introduction.}
Recent excitement in the literature has drawn attention to
quantum field simulators of cosmological evolution as implemented in cold atom systems
\cite{PhysT,science13chen,prx18Eckel,Neuenhahn_2015,
PRL08Berges,Berges2016,
PRL15Berges,
BERGES2002847,PRL12Berges,Eisert2015,pra21Chatrchyan,Oberthaler,Calzetta2005,Faccio2013}.
Such studies are motivated by the fact that
atomic physics laboratories provide the possibility of studying in real time
and in a reproducible fashion, the analogue of extreme non-equilibrium~
\cite{Keldysh:1964ud,schwinger,Perel,Kadanoff,pre94scrdnicki,mbl,nature17Moessner}
conditions such as might have prevailed in the early universe. Much of the emphasis, both in experiment~\cite{prx18Eckel,PhysT,Oberthaler}
and theory~\cite{Fischer1,Fischer2,Fischer3} has focused on the inflation stage.
Equally important are the three subsequent stages which have received some
attention
~\cite{pra21Chatrchyan,prx18Eckel} as well.

\begin{figure}
  \includegraphics[width=3.7in,clip]{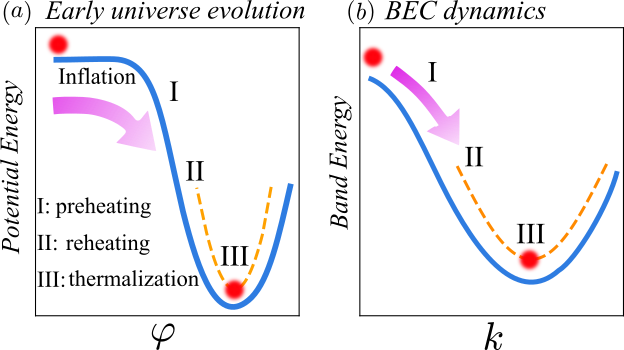}
  \caption{(Color online) Similarity of systems which transition from an unstable
to a thermal state. (a). Slow roll inflation picture reflecting how the inflaton slowly
rolls down the potential and ultimately generates
the standard model particles. (b). A Bose-Einstein condensate initially at an unstable state
similiarly rolls down to relax into the true ground state, corresponding to the band minimum. }
  \label{Cartoon}
\end{figure}

A paradigmatic cosmological example which describes the evolution
of the early universe
is the ``slow roll inflation" scenario~\cite{senatore2016,forconi2021,allahverdi2010,baumann2009,albrecht1982,kofman1994}.
The crucial features of this scenario are that (1) the universe is an isolated quantum system in which
inflationary processes proceed on their own.
(2) This scenario begins with a homogeneous scalar inflaton field $\varphi$
having high energy; the
subsequent dynamics correspond to
$\varphi$ slowly rolling down a potential energy hill towards equilibration.
After an exponentially slow inflation period at the beginning, (3) the inflaton oscillates and transfers its energy to matter fields involving an explosive particle production,
and then finally to thermalization. This cosmological model, importantly,
has a body of experimental support~\cite{senatore2016,forconi2021}.
These processes are illustrated schematically in the simple picture of Fig.~\ref{Cartoon}a,

In this paper we capture these three essential features of cosmology
(after inflation) by studying
the dynamics of a dilute Bose gas on an optical lattice,
transitioning from an unstable to a stable Bose-Einstein condensate (BEC).
Our unstable BEC is formed in an optical
lattice configuration through a quench that instantaneously pumps particles from a lower to an upper band.
Unlike the other atomic physics platforms~\cite{pra21Chatrchyan,prx18Eckel,Oberthaler}
ours is an isolated system which starts from a homogeneous BEC. The evolutionary dynamics proceeds on its own without external drive
so that all the dynamics we observe after the quench is driven by the excited BEC or
``inflaton".

We find that
this high-energy BEC state which is intrinsically unstable, spontaneously
transfers its energy to inhomogenous quantum fluctuations and finally
transitions to a stable (condensate) state.
The dynamical path is described by three evolutionary stages with striking similarity to the three post-inflation
stages of inflaton dynamics. Its subsequent evolution corresponds qualitatively to the simple dynamics suggested in
Fig.~\ref{Cartoon}b.

By exploiting this
correspondence, our
paper
demonstrates how cold atom systems can provide
\textit{concrete}, analytically tractable models for the dynamics within each of these evolutionary
stages of the early universe.
Accompanying our analytics are demonstrably consistent Gross Pitaevskii simulations which
establish the relative time intervals associated with each of the three stages.
Moreover, the thermalized state or endpoint contemplated here is a recondensation in contrast to previously studied
cosmological analogue systems.
Thus, this work should satisfy
a centrally important objective of cold atom research: to
create and to more deeply understand the pathways associated with the non-adiabatic preparation of
exotic condensate phases~\cite{Mueller,wang2021,fu2022}.

\begin{figure}[h]
  \includegraphics[width=.5\textwidth]{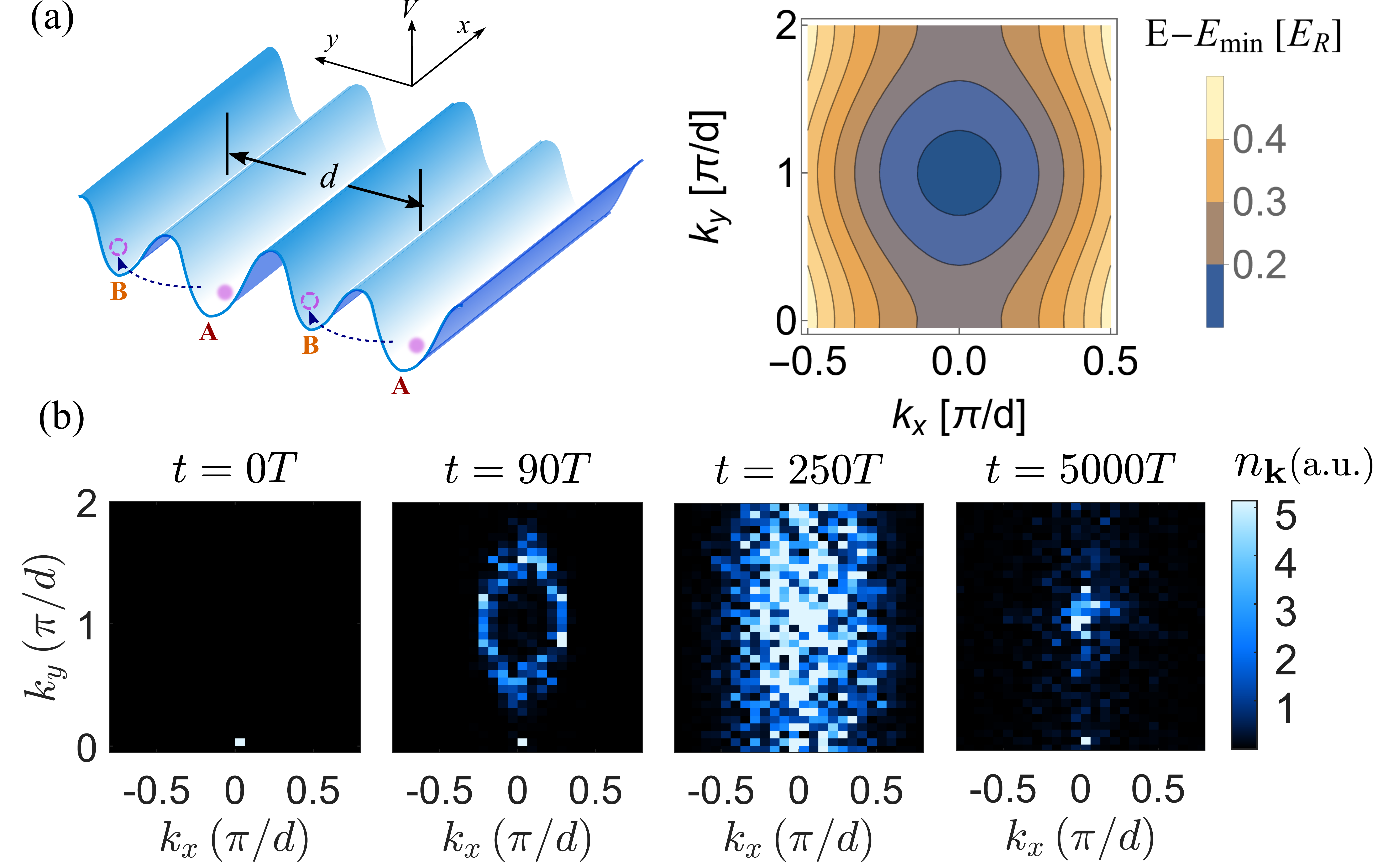}
  \caption{ (Color online)  Optical lattice set-up and numerical results. (a). Left panel: schematic plot of the 2D optical lattice potential confining atoms. The unit cell consists of two sites: A and B.
Pink circles in (a) denote atoms. Right panel: Color contour of the upper band structure. Here $k=0$ is a saddle point in 2D and a local maximum along the $k_y$ direction. (b). Momentum-space atom distribution $n_{\bf{k}}=|\psi({\bf{k}})|^2$ at different stages from GP simulations at given interaction strength. Here $\psi({\bf{k}})$ is the Fourier transform of $\psi({\bf{r}})$, Here the units of time are in recoil energy, $E_R$; for convenience in plotting we take $T$ to be ${h/(4 E_R)}$.
For numerical
details see Ref.~\onlinecite{Supplement}.}
  \label{schematic}
\end{figure}

Fig.~\ref{schematic}(a) provides an illustration of the proposed experiment.
The
momentum distributions of the immediate post-quench stage and that of the three subsequent
evolutionary stages obtained from our simulations are represented by the snapshots in Figure~\ref{schematic}b.
The first of the latter three
(corresponding to ``preheating" in the cosmological literature)
yields, through a parametric resonance, what we refer to as a ``ring condensate".
\footnote{This parametric resonance, in contrast to Ref. \onlinecite{pra21Chatrchyan}, is
naturally associated with the intrinsic dynamics of the isolated system and does not require an externally
imposed time dependent interaction,
as discussed in the Supplement.}
More precisely this so-called ``condensate" corresponds to a macroscopic occupation of
many finite momentum modes (as in a fragmented condensate) but
distributed in a ring geometry, as seen in the second panel of
Figure~\ref{schematic}(b).

Collisions between the initial and ring condensates lead to a second stage (``reheating") which ends with
a destruction of both condensates and a proliferation of non-thermal bosonic quasi-particles having a range of different momenta,
which is seen in
the third panel of Figure~\ref{schematic}(b)
and which we refer to as a {\it cloud state}.
The formation of this cloud, in which all phase coherence is lost is, in turn,
a crucial step that then enables the system to
reach the third stage corresponding to full thermalization shown
in the fourth panel
Figure~\ref{schematic}(b).

{\em Numerical simulation. }
As is illustrated in Fig.~\ref{schematic}a,
we start with a two-dimensional system where the Bose atoms are confined in the $y-$direction by a periodic
potential
$V(y)=V_1 \sin(4\pi y/d)+V_2 \sin(2\pi y/d)$
with a free particle dispersion in the $x-$ direction. Here $V(y)$ involves two sublattices (denoted as $A/B$ sites) with a potential offset controlled by $V_2$.
The potentials of A and B are exchanged during the quench, which effectively pumps atoms from A sites to B sites,
thereby exciting the BEC from the lower band to the upper band.
The extra free dimension in Fig.~\ref{schematic}(a) plays an essential role~\cite{Ketterle,fu2022} in the recondensation process, as it provides high energy states needed to absorb the released kinetic energy when the BEC reforms in the end. The results here can be readily extended to 3D systems with only quantitative changes.


%
In our simulations, we use a CUDA-based Gross-Pitaevskii equation solver implemented on graphic processing units, based on
a split-step algorithm \cite{Andreas,Kathy} with no dissipation added:
$i\hbar  \frac{\partial}{\partial t} \Psi(\mathbf{r},t)=\Big(-\frac{\hbar^2{\bf{\nabla}}^2}{2m} +{V(y)} +g|\Psi({\bf{r}},t)|^2\Big) \Psi(\mathbf{r},t)$, where $m$ is the boson mass, and $g$ the interaction strength.

{To make contact with cosmology we note that in the cold atom laboratory, a quantum quench of the BEC to the {upper} band is the analogue
creation of a homogeneous oscillating inflaton field. The quenched BEC is associated with a finite {oscillation} frequency $\omega\sim J/\hbar $,
where $J$ corresponds to the {width of the upper band with} dispersion $\epsilon_{\bf{k}}=\hbar^2k_x^2/2m+|J| \cos(k_y d)$. The subsequent dynamics is essentially confined to the upper band since
the interaction-mediated tunneling between two-bands is {tuned} to be negligibly small.
We presume $gn_0/J < 1$ {(where $n_0$ is the condensate density)}
and from our tight-binding limit simulations determine the value for $J \approx 0.05 E_R$ in recoil units.
It is this finite-frequency BEC which serves as an internal driving source to pump particles out of the condensate,
leading to its fragmentation and eventual disappearance.}

{\em Preheating: early dynamics.}
{In cosmological models, at the end of the inflation period the inflaton field
is assumed to oscillate around the minimum of its potential and in this way decay into other forms of matter. This
next stage following inflation is called preheating\cite{ABBOTT198229,PhysRevD.42.2491,Felder2001},
where the universe is populated via parametric resonances.
The coherent nature of the BEC-inflaton enhances the efficiency of particle production
and
at early times particles (inhomogeneous fluctuations) are generated exponentially.}

Analogously, these parametric resonances emerge in the cold atom system during the preheating stage which is associated with
an effective preheating Hamiltonian~\cite{inflaton}
\begin{eqnarray}
\label{forming_ring}
\hat{H}^p_{\text{eff}}=\sum_{\mathbf{k}\neq 0}\epsilon'( \mathbf{k}) \hat a^\dagger_{ \mathbf{k}} \hat a_{ \mathbf{k}}   +\frac{g n_0  }{2 } \sum_{\mathbf{k}\neq \mathbf{0}}  ( \hat a^\dagger_{\mathbf{k}}  \hat a^\dagger_{-\mathbf{k}}+ \hat a_{\mathbf{k}}  \hat a_{-\mathbf{k}}  ).
\end{eqnarray}
Here  $\epsilon'({\bf k})=\epsilon_\mathbf{k}-J+ {gn_0}$, ${\hat a_{\mathbf{k}}}$ is a bosonic operator.
The presence of imaginary eigenvalues $\lambda({\bf k})\equiv \sqrt{ \epsilon'( \mathbf{k}) ^2-g^2n_0^2}$ of Eq.~\ref{forming_ring}
(when $|\epsilon'( \mathbf{k})|<gn_0$)
reflects exponential growth of the new condensate particles. The resonance condition $\epsilon_\mathbf{k}=J- gn_0$ corresponds,
in momentum space to a ring shaped additional fragmented condensate, which forms
on a time scale
$t_r\sim\hbar (gn_0)^{-1}\ln L$. Here $L$ denotes the system size.
These predictions are confirmed by the GP simulations shown in
the second panel of Fig.\ref{schematic}(b).  The dynamics can be described
by an analogue of the Mathieu equation~\cite{Supplement}.

\begin{figure}[h]
  \includegraphics[width=.5\textwidth]{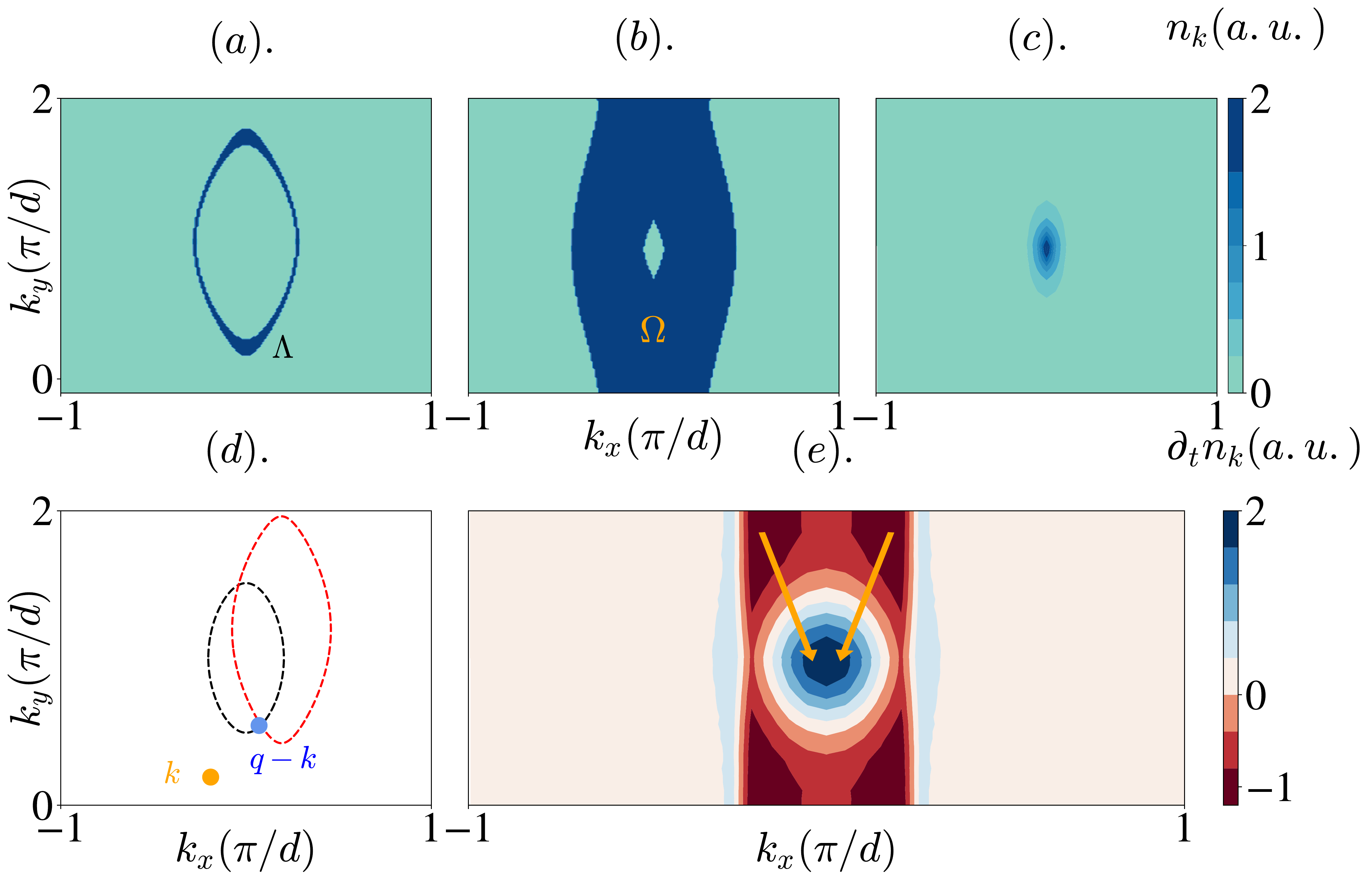}
  \caption{ Evolutionary stages derived from analytic theory (a)-(c) which
should be compared with simulations in Fig.~\ref{schematic}(b). Here, the ring stage (a) originates from a parametric instability
while (b) the cloud stage arises from collisions between two condensates. The
quasi-thermal distribution (c) derives from Boltzmann dynamics. (d). Characterization of the
${\bf{k}}$ and ${\bf{q}}-\bf{k}$ pairs which lead to cloud formation and derive from momentum and energy conservation~\cite{footnote1}
(e). Distribution
of the rate of change of particle number $\partial_t n_{\bf{k}}$. The plot presents
this distribution at early times where Boltzmann dynamics is applicable. The arrows show how particles flow to the band minima.
}
\label{fig:2}
\end{figure}

{\em Reheating stage. } 
The universe enters the reheating stage once the energy of the newly-populated degrees of freedom becomes non-negligible. Here the inflaton continues to oscillate and interact with its fragmented products (called re-scattering); this is associated with
non-linear dynamics. At the same time the energy of the inflaton and its by-products
is transferred to standard model particles.
The accompanying phenomenology of this process in cosmological models is
complex. There may emerge turbulent scaling\cite{PhysRevD.54.3745}, or
oscillons\cite{PhysRevD.90.083528,PhysRevLett.108.241302} or solitons\cite{TKACHEV1998262} and cosmic defects. 

In the cold atom set-up, the formation of the ring-condensate in momentum space shown in Fig.~\ref{fig:2}(a)
marks the start of the reheating stage. In this stage, the quenched BEC (inflaton)
continues to oscillate at high frequency and interactions between the BEC and the ring-products become important. Particles are pumped out of these
by-products and energy is transferred to other non-condensed particles. These
dynamical processes eventually generate a highly non-equilibrium ``cloud"
phase shown in the third panel of Fig.~\ref{schematic}b and in ~\ref{fig:2}b.

It is convenient at this point to introduce a characterization of the momentum regime occupied by the ring condensate;
we call this $\Lambda$ throughout the paper (see Fig.~\ref{fig:2}(a)).
One important interaction
effect \cite{Supplement}, $V_{ \Lambda,\Lambda}$ corresponds to the annihilation of two particles in the ring condensate. These collision events are associated with back-reaction dynamics and broaden the radius of the ring.
More important is the second process
$V_{  0,\Lambda}$ which annihilates particles from both condensates and
leads to
the destruction of condensates associated with an effective reheating Hamiltonian
~\cite{Supplement}:
\begin{eqnarray}
\label{5}
\hat{H}^{\bf 0,r}_{\text{eff}}&\equiv&\sum_{ \mathbf{q}\notin \Lambda }  \epsilon''_\mathbf{q}  \hat a^\dagger_{ \mathbf{q}} \hat a_{ \mathbf{q} } +
U'  \sum_{\mathbf{k}\in \Lambda,\mathbf{q}\notin \Lambda } ( {\hat a}_{\mathbf{q}}^\dagger \hat a^\dagger_{\mathbf{k}-\mathbf{q}}+h.c.  ) \nonumber \\
&+&U'  \sum_{\mathbf{k}\in \Lambda,\mathbf{q}\notin \Lambda } (  {\hat a}_{\mathbf{q}}^\dagger \hat a_{\mathbf{q}-\mathbf{k}}+h.c. ).
\end{eqnarray}
Here $\epsilon''_\mathbf{q}=\epsilon_\mathbf{q}-(J-gn_0/2+  g\bar{n}_r/2
-g\bar{n}_0/2)$ is an effective kinetic energy and $U'=2g\sqrt{ \bar{n}_0 \bar{n}_r/L_r }$
is the effective interaction strength for these back-reaction processes with $L_r$ being the number of modes in the ring. Additionally, $\bar{n}_{0,r}$ is the respective density for each co-existing condensate.

In contrast to
Eq.~\ref{forming_ring}, where the physics is local in $\boldsymbol{k}$-space,
the Hamiltonian in Eq.~\ref{5} is intrinsically non-local.
These
non-local features, reflecting the extended ring condensate,
are generic and universal and pertain to a
geometrically extended ``resonance band" in  $\boldsymbol{k}$-space.
The physical consequences of Eq.~\ref{5} are that the eigenvalue spectrum now involves
a large number of
{\it complex} values, $\lambda_i$ where the range of $i$ scales with the system size.
These complex eigenvalues suggest an interpretation in which
there is a proliferation of bosonic particles concurrent with the decay of the condensate(s).

\begin{figure}[h]
 \includegraphics[width=.5\textwidth]
{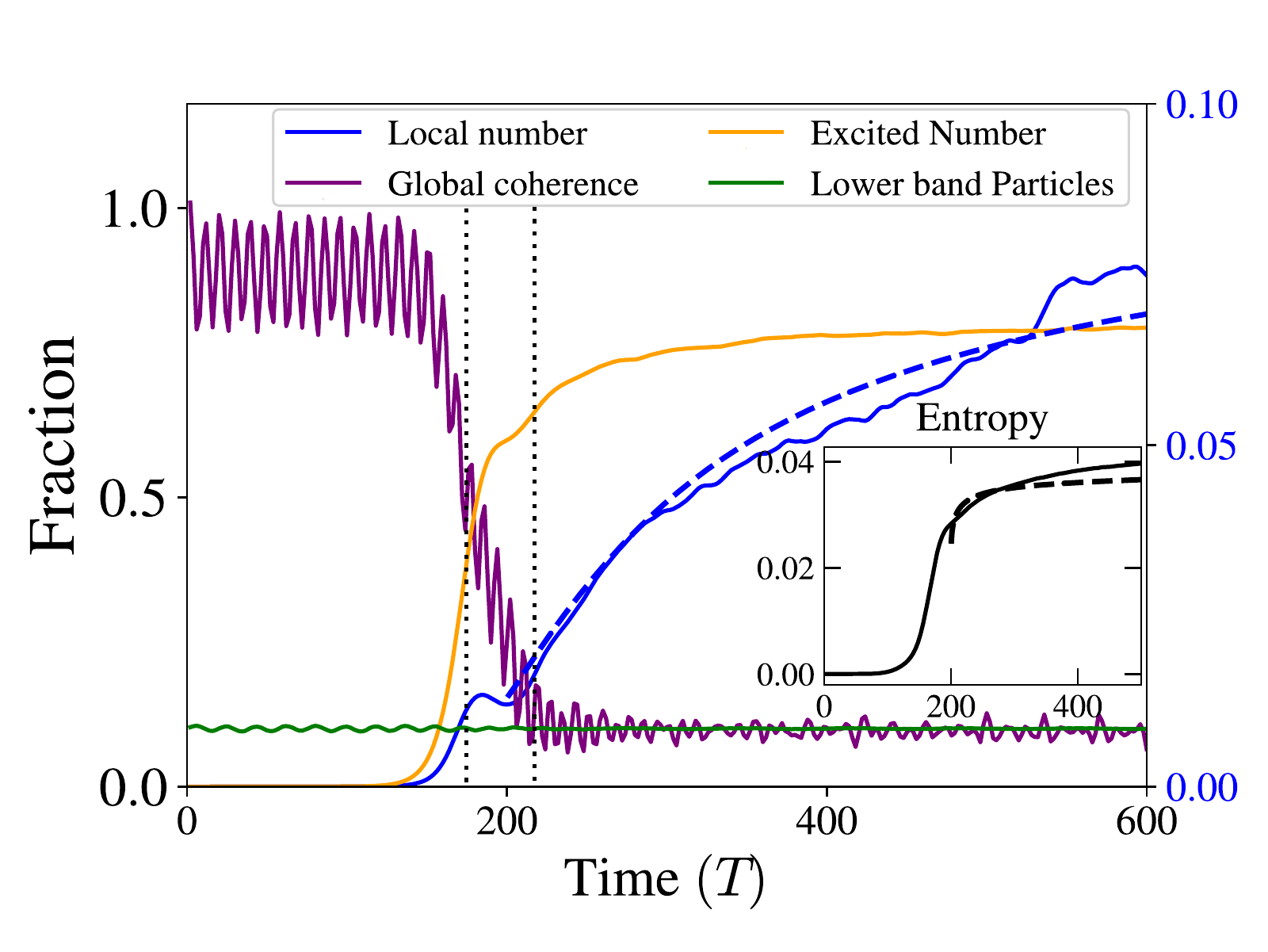}
  \caption{ Time evolution of key properties from GP simulations.
Beyond $600T$ the behavior is rather stable.
The $y$-axis for blue lines is on the right,
while the left axis pertains to all other curves.
The solid and dashed blue lines compare the particle number $n_\mathbf{k}$ near the band minima
from GP and analytical Boltzmann calculations, respectively.
The purple curve reflects global phase
coherence while the
orange line plots the total particle number
in excited states excluding the $\mathbf{k}=0$ mode.
The green curve indicates the residual particle number in lower band, and is responsible for
the oscillatory behavior found in the global coherence plots.
Vertical black dotted lines indicate the
time interval over which there is a smooth crossover between different evolutionary stages. The inset compares the entropy per particle from GP simulations (solid) and Boltmann theory (dashed) using  $S=N^{-1}
   \sum_{\bf{k}}\left[(1+n_{\bf{k}})\log(1+n_{\bf{k}})-n_{\bf{k}}\log(n_{\bf{k}})\right]$, where $N$ is the total particle number.  This monotonically increasing entropy and its final saturation
at long times are consistent with equilibration.
}
  \label{fig:4}
\end{figure}

To understand the origin of these complex eigenvalues and the time evolution more
quantitatively, we note that
the dynamics associated with
Eq.~\ref{5}
can be derived using a $2\times 2$ matrix; since the effective interaction $U'$ is small, each pair $(\mathbf{k}, \mathbf{q}-\mathbf{k})$
can be treated separately. This corresponds to an equation of motion
\begin{eqnarray}
\label{Eom}
\frac{d}{dt} \begin{pmatrix}
\hat a_\mathbf{k}(t) \\
{\hat a}^\dagger_{\mathbf{q}-\mathbf{k}}(t)
\end{pmatrix}\simeq \frac{- i}{\hbar}\begin{pmatrix}
\epsilon''_\mathbf{k} & U'     \\
- U' & -\epsilon''_{\mathbf{q}-\mathbf{k}}
\end{pmatrix}\begin{pmatrix}
\hat a_\mathbf{k}(t) \\
{\hat a}^\dagger_{\mathbf{q}-\mathbf{k}}(t)
\end{pmatrix}
\end{eqnarray}
where ${\hat a_\mathbf{k}(t)}$ evolves under the effective Hamiltonian in Eq.~\ref{5}. The eigenvalues
are given by
\begin{equation}
\lambda_\pm(\mathbf{k},\mathbf{q}-\mathbf{k})=\frac{\epsilon''_\mathbf{k}-\epsilon''_{\mathbf{q}-\mathbf{k}}}{2}\pm \sqrt{
\left( \frac{\epsilon''_\mathbf{k}+\epsilon''_{\mathbf{q}-\mathbf{k}}}{2} \right)^2
-{U'}^2 }
\end{equation}
and when $\epsilon''_\mathbf{k}+\epsilon''_{\mathbf{q}-\mathbf{k}}=0$, one clearly sees that the corresponding values for
$\lambda$ become {\it complex}.

The momentum distribution ($\Omega$) associated with this reheating stage is plotted in Fig.~\ref{fig:2}(b)
which takes on a cloud-like form representing a proliferation of non-condensed, non-thermal bosons.  { Here
the system enters into a highly-non-equilbrium phase, where the initial condensate, ring-condensate and a cloud-shaped
distribution of non-thermal bosons co-exist. }
This cloud, deriving from the complex eigenvalues in Eq.~\ref{5} reflects collisions between specific bosonic pairs as
shown in Fig.~\ref{fig:2}(d).
Here the energy and momentum
conservation constraints, indicated by the two dashed lines, establish how to associate the value of $q-k$ with a given $k$ in the pair. The reheating stage ends
finally with the complete decay of the condensate and the full formation of the cloud state.

This condensate decay
is the cold gas counterpart of the cosmological analogue in which at the end of reheating one has
the generation of
standard model particles. Here one sees
a destruction of all vestiges of phase coherence observed in the earlier evolutionary stages,
in many ways similar to the cosmological picture in which
the original memory of the (inflaton-)condensate completely disappears\cite{Supplement}.

{\em Late time: thermalization.}
In cosmology, the reheating stage ends with the complete decay of the inflaton field and a highly non-thermal distribution of standard model particles.
All important particles which will ultimately evolve to thermal equilibrium are already generated.
What follows next
is a stage of thermalization in which energy is redistributed
among the particles; this
enables them to reach an
equilibrium distribution, as supported by cosmological evidence\cite{cos1}.

In a similar way in the cold atom system, the quenched condensate completely disappears and a highly non-thermal distribution of particles emerges, marking the end of the reheating stage.
Thermalization then follows; here all finite-momentum degrees of
freedom {equilibrate} as the system relaxes towards thermal equilibrium.  We find that the dynamics in this stage is well described by a quantum Boltzmann equation.
This evolves the non-thermal cloud state to a quantum Bose-Einstein distribution
with, as it turns out, nearly 20\% of particles in the condensate.

In this Boltzmann-like description of the dynamics the interaction energy is assumed to be much smaller than the bandwidth so that it can be treated perturbatively. This leads to the famous quantum kinetic Boltzmann equation for a Bose gas\cite{Zakharov1992},
\begin{eqnarray}
\label{8}
\partial_t  n_{\mathbf{k}} =  I\left(\{ n_{\mathbf{k}'} \}\right).
\end{eqnarray}
Here $n_{\mathbf{k}}$ is the particle number distribution function and $I\left(\{ n_{\mathbf{k}'} \}\right)$ is the collision integral. Perturbatively we have that $I\left(\{ n_{\mathbf{k}'} \}\right)$ depends on: \begin{eqnarray}
\label{9}
\Gamma(\mathbf{k},\mathbf{k}';\mathbf{k}+\mathbf{q},&\mathbf{k'}-\mathbf{q})=n_{\mathbf{k}-\mathbf{q}} n_{\mathbf{k}'+\mathbf{q}} (1+n_\mathbf{k}) (1+n_{\mathbf{k}'}) \nonumber \\
&- (n_{\mathbf{k}-\mathbf{q}}+1) (n_{\mathbf{k}'+\mathbf{q}}+1)  n_\mathbf{k}  n_{\mathbf{k}'}  .\end{eqnarray}
Here $\mathbf{k}'$ and $\mathbf{q}$ are integrated out to yield the collision term, $I= 2\hbar^{-1}g^2  \int d^2 \mathbf{k}' d^2\mathbf{q}(2\pi)^{-3}\Gamma \delta(E_i-E_f)$, where for simplicity of notation we have dropped the arguments in $\Gamma$; the $\delta(E_i-E_f)$ term
introduces conservation of kinetic energy. This collision integral then determines the momentum and energy flow processes which lead the system to
equilibration.

Because of the special property of the cloud state, initially, the main contribution to $\Gamma$ is from scattering events involving one out-of-cloud and three in-cloud modes \cite{Supplement}. In such events, $\Gamma$ is dominated by the cubic terms \cite{Zakharov1992} $\sim n^3$.
More specifically the important collisions derive
from two intermediate-energy particles in the cloud which scatter into one low-energy mode
around the band minimum and the other mode at high energy.
The flow of particles is shown in Fig.~\ref{fig:2}(e).
As a result, Eq.~\ref{8} is of the form
$\partial_t n_\mathbf{k} \propto {g^2} n^3_0$.
As a consequence
of the relatively time independent particle distribution, it follows that
the initial dynamics evolving from the cloud yields a linear-in-time growth for the
occupation of the band minimum.

As the dynamics continues to evolve in time the system will eventually be driven towards a Bose-Einstein
distribution. In this regime, the
quadratic term $n_{\mathbf{k}-\mathbf{q}} n_{\mathbf{k}'+\mathbf{q}}-n_{\mathbf{k}} n_{\mathbf{k}'}$ in $\Gamma$
begins to dominate and
the time variation of the particle distribution slows down, approaching
the quantum distribution function $n^{\text{eq}}_{\mathbf{k}}=\left( e^{\beta (\epsilon_\mathbf{k}-\mu)} -1\right)^{-1}$, as expected.
Here the inverse temperature $\beta$ in this equilibrium state is determined by the kinetic energy of the initial cloud while
the chemical potential $\mu$ approaches zero for sufficiently large system sizes. In this way
we establish {\it condensation} at the band-minima in the thermodynamical limit.

\textit{Comparison between analytics and simulations-}
We turn now to Figure \ref{fig:4} which summarizes the evolutionary stages as found in the GP simulations
and presents comparisons with our analytics, demonstrating reasonable consistency.
There are multiple time dependent functions indicated by the curves.
At early times
(of the order of $t_1\propto g^{-1}$)
the simulations show that
there is a rapid growth in occupation of excited modes (orange curve) which reflects the formation of a new
ring-like condensate. The governing dynamics shows
an exponential growth rate, as expected in a parametric resonance.
Following this,
in the second stage, the system experiences a complete loss of global phase coherence
(purple curve) associated with the upper band, which corresponds to the cloud stage.
Following this stage, then, the occupation number in the band minimum (solid blue curve)
grows appreciably, displaying the expected linear time dependence for an extended range of
intermediate times.

The number of particles at the band minimum is shown
in Fig.~\ref{fig:4}
via a comparison between simulations (solid) and Boltzmann analytical calculations (dashed)
blue curves,
indicating that both overlap reasonably well.
This stage of thermalization is reflected
in the growth of the (H-theorem) entropy as well, which is plotted as analogous dashed and solid
lines in
the inset to Figure \ref{fig:4}.

\textit{Conclusions-}
{
This paper has shown how
the intrinsic dynamics of a Bose condensate in an optical lattice transitioning from an unstable to
a stable BEC
provides an analogue experiment
for the slow roll inflation cosmology
scenario.
Importantly, it represents a homogeneous and a closed quantum system in which the inflationary processes
proceed without the external drive which is usually incorporated in these analogue laboratories.
In this way it provides a complement to other
exciting work
\cite{PhysT,science13chen,prx18Eckel,Neuenhahn_2015,
PRL08Berges,Berges2016,
PRL15Berges,
BERGES2002847,PRL12Berges,Eisert2015,pra21Chatrchyan,Oberthaler,Calzetta2005,Faccio2013}
addressing
cosmology as seen through the lens of
a quantum gas.
This fairly generic equilibration procedure informs about cold atom engineering of exotic condensate phases
which should be of direct interest to the cold atom community. It, moreover,
leads to multiple testable predictions.
Equally important is that such studies provide
analytically tractable models for the dynamics within each of these evolutionary
stages.
}

\vskip2mm

 {\em Acknowledgements-}
We thank Cheng Chin for his substantial help in writing this manuscript.
Additionally we are very grateful to Erich Mueller for
calling attention to the specific quench protocol studied here.
This work was partially (KL, KW) supported by the Department of Energy
(DE-SC0019216).
HF acknowledges support from DOE, Grant No DE-SC0022245.
We thank Andreas Glatz for valuable help and insight over many years of
collaborations.
We also acknowledge Steven Wu, D. T. Son, L. T. Wang, S. Koren, A. Lucas, Y. M.
Zhong and  G. H Zhou for helpful discussions.

\bibliography{document}

\end{document}


\widetext

\begin{center}
\large{\textbf{Supplement: Simulating Cosmological Evolution by Quantum Quench of an Atomic BEC }}
\end{center}

\beginsupplement

In this supplement we present additional analytical details and simulation results on
the evolutionary dynamics of quenched Bose-Einstein condensates (BEC). We analyze each stage sequentially in Secs. I-III, emphasizing analogies with
cosmological evolution. Also, we discuss some analytical results deduced
directly from the Gross-Pitaevskii (GP) equation, thereby providing additional insights into
our numerical simulations. In the last section, we include details of numerical parameters used for our simulations.

\section{Dynamics in the preheating stage}
\vskip2mm

To begin, we address the early dynamics in the preheating stage, focusing on an effective theory of the upper band degrees of freedom, with the effective Hamiltonian
\begin{eqnarray}
\label{upper}
H_{\text{upper}}= \sum_{\bf{k}}  \epsilon_{\bf{k}} \hat{a}^\dagger_{\bf k} \hat{a}_{\bf k}+\frac{g}{2V}\sum_{\bf{k},\bf{k'},\bf{q}}  {\hat a^\dagger_{\bf{k}} {\hat a}^\dagger_{\bf{k}'}\hat a_{\bf{k}'-\bf{q}}\hat a_{\bf{k}+\bf{q}}}.
\end{eqnarray}
Here $\epsilon_{\bf{k}}=\hbar^2k_x^2/2m+|J| \cos(k_y d)$  is the upper band dispersion and  {$\hat a_\mathbf{k},\,\hat a_\mathbf{k}^\dagger$ are}  bosonic operators in this band, $J$ is the hopping parameter associated with the periodic potential
treated here in a tight-binding approximation. To obtain $J$, we directly diagonalize the original Hamiltonian matrix based on the implemented lattice potential with a size cutoff where a convergence is achieved. $J$ is exponentially dependent on the lattice depth (dominated by $V_1$ here) while the band gap between first and second bands are tuned by $V_2$.  Because of the macroscopic occupation of a pumped BEC, we focus on
the dynamics of scattering events associated with $\mathbf{k}=\mathbf{0}$ modes. This leads to
a many-body interaction term,
\begin{eqnarray}
\hat V_0\equiv \frac{g}{2V} \Big(
 {\hat a^\dagger_\mathbf{0} } \hat a^\dagger_{\mathbf{0}}  \hat a_{\mathbf{0}}   \hat a_{\mathbf{0}}+  \sum_{\mathbf{k}} ( \hat
a^\dagger_\mathbf{k}  \hat a^\dagger_{\mathbf{-k}}  \hat a_{\mathbf{0}}   \hat a_{\mathbf{0}}+h.c.  )+ 4\sum_{\mathbf{k}} \hat
a^\dagger_\mathbf{k}  \hat a^\dagger_{\mathbf{0}}  \hat a_{\mathbf{k}}   \hat a_{\mathbf{0}}.
\Big)
\end{eqnarray}


It is useful to rewrite this expression using $\hat a^\dagger_\mathbf{0} \hat a_\mathbf{0}=N- \sum_{\mathbf{k}\neq \mathbf{0}} \hat a^\dagger_\mathbf{k} \hat a_\mathbf{k}$ where the scalar $N$ is the total number of particles in the system.
Using this replacement, one finds a quadratic effective Hamiltonian describing the scattering events between condensed and non-condensed particles {\cite{inflaton}},
\begin{eqnarray}
\label{S3}
\hat{H}^p_{\text{eff}}=\sum_{\mathbf{k}\neq 0} \epsilon'_\mathbf{k}\hat a^\dagger_{ \mathbf{k}} \hat a_{ \mathbf{k}}   +\frac{g n_0  }{2 } \sum_{\mathbf{k}\neq \mathbf{k}}  ( \hat a^\dagger_{\mathbf{k}}  \hat a^\dagger_{-\mathbf{k}}+ \hat a_{\mathbf{k}}  \hat a_{-\mathbf{k}}  ).
\end{eqnarray}
Here $\epsilon'_\mathbf{k}=\epsilon_\mathbf{k}-J+ {gn_0}$ {, $n_0=N/V$}, and we have ignored all quantum fluctuations from non-condensed particles. This yields Eq.~3 in the main-text. In the Heisenberg picture,
we address the time evolution of operators under this effective Hamiltonian,
 \begin{eqnarray}
 \hat a_{\mathbf{k}}(t)=e^{i \hat{H}^p_{\text{eff}} t {/\hbar}}  \hat a_{\mathbf{k}} e^{-i \hat{H}^p_{\text{eff}} t {/\hbar}}
 \end{eqnarray}

From the commutation relations
it is seen that
\begin{eqnarray}
\frac{d}{dt} \begin{bmatrix}
 {\hat a_\mathbf{k}}(t) \\
  {\hat a^\dagger_{-\mathbf{k}}}(t)
\end{bmatrix}=   {\frac{-i}{\hbar}}\begin{bmatrix}
\epsilon'_\mathbf{k} &{g n_0}    \\
{ {-}g n_0}   & -\epsilon'_{\bf{k}}
\end{bmatrix}\begin{bmatrix}
 {\hat a_\mathbf{k}}(t) \\
 {\hat a^\dagger_{-\bf{k}}}(t)
\end{bmatrix}.
\end{eqnarray}
Note that the $2\times 2$ matrix here is non-hermitian so that the dynamics are non-unitary.

From the matrix eigenvalues,
\begin{eqnarray}
\label{S6}
\omega_\mathbf{k}=\sqrt{\epsilon'^2_\mathbf{k}-  {g^2 n^2_0} } {/\hbar} .
\end{eqnarray}
it follows that $\omega_\mathbf{k}$ becomes imaginary when $\epsilon'^2_\mathbf{k}-  {g^2 n^2_0}<0$
which is to be associated with exponential growth in the number of particles. The condition $\epsilon'_k=gn_0$ provides the boundary separating the dynamics associated with
that of conventional Bogoliubov quasi-particles and that associated with
a parametric instability leading to the growth of a ring
condensate, as discussed in the main text.

The ring center is determined by
\begin{equation}
\label{eq:ring}
\hbar^2k_x^2/2m+J \cos(k_y d)=J-gn_0.
\end{equation}
Since $\cos (k_y {d})$ is periodic in momentum space, this leads to a closed ring.

Moreover, the boundary region for exponential growth corresponds to the onset of imaginary
eigenvalues; this is associated with the condition $|\epsilon'_k|=gn_0$. At this boundary
the modes are static, having zero frequency.
%

\begin{figure}[h]
  \includegraphics[width=.5\textwidth]{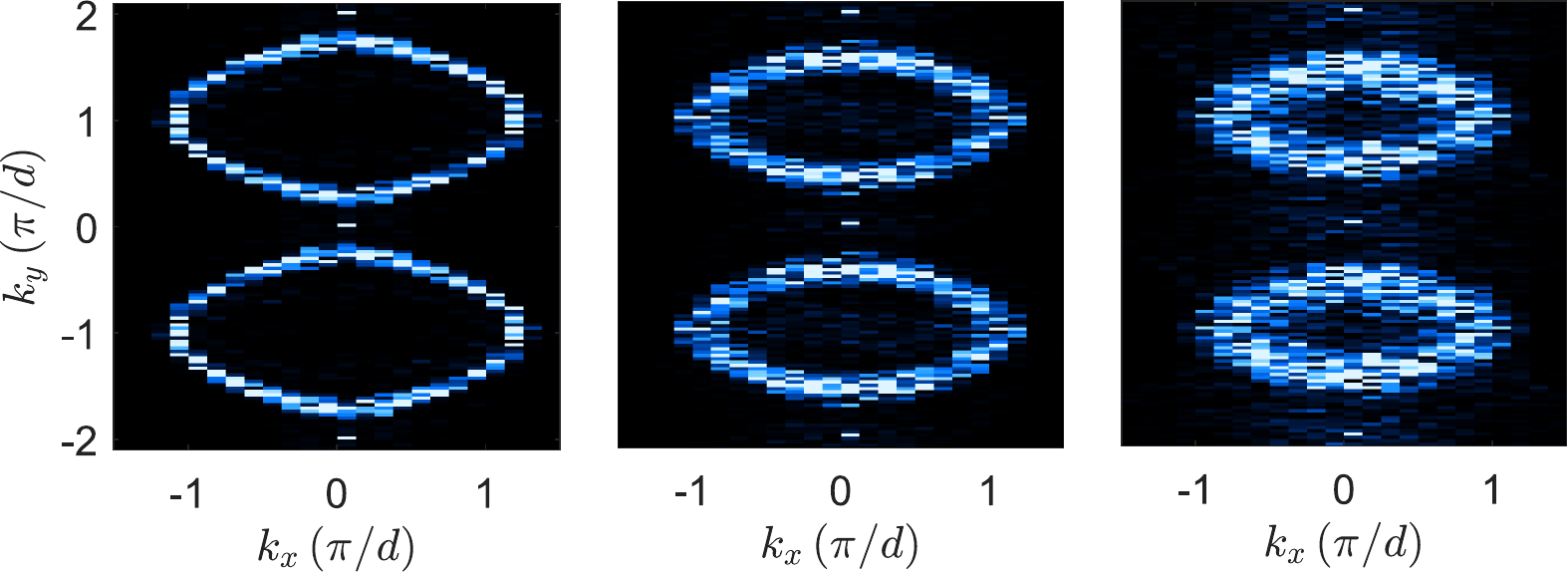}
  \caption{Ring-formation during the preheating stage, as represented by a momentum-space distribution.
 From left to right, the interaction energy is increased as $gn_0=0.0075E_R,\,0.0151E_R,\,0.0226E_R$.  {$n_{\bf{k}}=|\psi({\bf{k}})|^2$}.  A clear scaling of ring width with the interaction energy determined by the resonance condition is seen.
  }
  \label{fig:ring}
\end{figure}

\textit{Connection to GP simulation results---} From Eq.\eqref{eq:ring}, one sees that the ring position is directly determined by the interaction strength. This is confirmed by numerical simulations, see Fig.~\ref{fig:ring}.
A more quantitative analysis is summarized in Table~\ref{tab:1}, where a scaling of the ring position and growth exponent is shown.

\begin{table}[h]
\begin{tabular}{|c|c|c|c|}
  $gn_0/h$ (Hz)& $t\,(T)$& $k_y\,(\pi/d)$& $\Delta E\, (E_R)$\\ \hline
  10 &310 &0.28&  -0.0189\\ \hline
  20 & 160  &   0.44 & -0.0425 \\ \hline
  30 & 110 & 0.56&  -0.0620\\ \hline
  40 & 90 &0.75 & -0.0892\\
 \end{tabular}
  \caption{Summary of characteristic timescale and position for the ring formation at different interaction strengths. $E_R/h= 1.3$ kHz, $2\pi/T=6300$ Hz. Here $k_y$ is identified as the smallest $y$-momentum value in the ring, which corresponds to $k_x$=0 point in the equi-energy ring in Fig. \ref{fig:ring}. The time is chosen such that the particle density at this $k_y$ has grown to around ~1000 (a.u.). Here the energy difference $\Delta E=E(0,k_y)-E_0$ is measured with respect to the initial energy $E_0=-2.07237E_R$ at ${\bf{k}}=0$ in the second band. One can see a scaling of $1/t$ and $\Delta E$ as $\propto gn_0$. Considering the error of $\Delta t\sim10T$, $\Delta k_y\sim0.03 (\pi/d)$ in numerical data, the scaling relation is reasonably well satisfied.} \label{tab:1}
\end{table}

 \section{Dynamics in the reheating stage}

The appearance of a second condensate (through ring formation) is an important step en route to
thermalized recondensation. Crucial here is that now two condensates are present
which can lead to inter-condensate scattering events. The ring is not a replacement
of the original condensate but rather represents a fragmentation.

\subsection{Back-reaction: balancing between $k=0$ and ring condensate}
In this subsection which is associated in the cosmological context with
``back-reaction",
we study the scattering events within ring condensates, first
establishing a mechanism to enable a co-existing BEC (corresponding to simultaneous
ring and $k=0$ condensates).
Ultimately in the ``re-scattering" phase, collisions
between the two lead to a de-stablization.

Once a ring BEC has formed, one has to consider scattering events involved with ring modes.
Here there are four terms to consider:
 \begin{eqnarray}
\hat{V}_{r,r}&\equiv&\frac{g}{2V}\left(\sum_{\mathbf{q}_1,\mathbf{q}_2 \in\Lambda }  \sum_{\mathbf{q}_3,\mathbf{q}_4\in\Lambda}+\sum_{\mathbf{q}_1,\mathbf{q}_2 \in\Lambda }  \sum_{\mathbf{q}_3,\mathbf{q}_4\notin\Lambda}+\sum_{\mathbf{q}_1,\mathbf{q}_2 \notin \Lambda }  \sum_{\mathbf{q}_3,\mathbf{q}_4 \in\Lambda}+4\sum_{\mathbf{q}_1,\mathbf{q}_3 \in\Lambda }  \sum_{\mathbf{q}_2,\mathbf{q}_4\notin\Lambda} \right)\nonumber\\
&\times& {\hat a_{\mathbf{q}_4}^\dagger} \hat a^\dagger_{\mathbf{q}_3}\hat a_{\mathbf{q}_2} \hat a_{\mathbf{q}_1} \delta(\mathbf{q}_1+\mathbf{q}_2-\mathbf{q}_3-\mathbf{q}_4).
\end{eqnarray}
Since the ring is an extended object, the first summation itself contains two types of contributions: (1) forward scattering events (such as $\sum_{\mathbf{q}_1,\mathbf{q}_2\in\Lambda} 2 a_{\mathbf{q}_1}^\dagger \hat a^\dagger_{\mathbf{q}_2}\hat a_{\mathbf{q}_2} \hat a_{\mathbf{q}_1}$); (2) the scattering involving
two incident particles with opposite momenta ( $\sum_{\mathbf{q}_1,\mathbf{q}_2\in\Lambda}
a_{\mathbf{q}_1}^\dagger \hat a^\dagger_{-\mathbf{q}_1}\hat a_{\mathbf{q}_2} \hat a_{-\mathbf{q}_2})$. In the second/third summation, the case $\mathbf{q}_1=-\mathbf{q}_2$ is the leading contribution while all other are sub-leading
\footnote{To observe why $\mathbf{q}_1=-\mathbf{q}_2$ is the leading contribution, one may define $f(Q)$ as the multiplicity for the space $q_1+q_2=Q$,  {i.e.,} $f(Q)=\text{Number of elements in }\{ (q_1,q_2) | \, q_1+q_2=Q,\, q_{1,2} \in \Lambda \}.$ One can check that, for the ring-shape $\Lambda$, $f(Q)$ is  {on} the order of $1$ if $Q\neq 0$ while for $Q=0$,  {$f(Q)$ is of the order $L\gg 1$ }. }. In the fourth summation, the leading contribution corresponds to $\mathbf{q}_1=\mathbf{q}_3$.

It follows that $\hat V_{r,r}$ can be approximated by
\begin{eqnarray}
\hat{V}_{r,r}&\simeq&\frac{g}{2V} \Bigg(\sum_{\mathbf{q}_1,\mathbf{q}_2\in\Lambda} (2 \hat a_{\mathbf{q}_1}^\dagger \hat a^\dagger_{\mathbf{q}_2}\hat a_{\mathbf{q}_2} \hat a_{\mathbf{q}_1}+ \hat a_{\mathbf{q}_1}^\dagger \hat a^\dagger_{-\mathbf{q}_1}\hat a_{\mathbf{q}_2} \hat a_{-\mathbf{q}_2})\nonumber \\
&+&\sum_{\mathbf{q}\in \Lambda,\mathbf{q'}\notin \Lambda }(\hat a^\dagger_{\mathbf{q}}\hat a^\dagger_{-\mathbf{q}}\hat a_{\mathbf{q'}}\hat a_{-\mathbf{q'}} + h.c. )+4\sum_{\mathbf{q}\in \Lambda,\mathbf{q'}\notin \Lambda }\hat a^\dagger_{\mathbf{q}}\hat a^\dagger_{\mathbf{q}'} \hat a_{\mathbf{q'}}\hat a_{\mathbf{q}}  \Bigg).
\end{eqnarray}

Again one can apply a mean field (M. F.) approximation and transform ${\hat{V}}_{r,r}$ to quadratic form. One may replace  {$\langle {\hat a^\dagger}_{\bf{q}}\hat a_{\bf{q}} \rangle = N_r/L_r$ and $\langle\hat a_{\bf{q}} \hat a_{-\bf{q}} \rangle = -i N_r/L_r$ if ${\bf{q}}\in\Lambda$.} Here  {$L_r$} counts the allowed momentum modes in the ring and $N_r$ is number of particles inside the ring. 
As discussed earlier for the preheating stage, one thus obtains an effective Hamiltonian containing
scattering events between particles on the ring condensate, and all other modes,
among ring condensed modes and other modes,
\begin{eqnarray}
\label{8}
H^r_{\text{eff}}=\sum_{ \mathbf{k}\notin \Lambda } (\epsilon_k- J+gn_0- g {\bar {n}_r}) \hat a^\dagger_{ \mathbf{k}} \hat a_{ \mathbf{k}} + i g {\bar{n}_r}(\hat a_{ \mathbf{k}} \hat a_{ -\mathbf{k}} - \hat a^\dagger_{ \mathbf{k}} \hat a^\dagger_{ -\mathbf{k}}),
\end{eqnarray}
where  {$\bar{n}_r=N_r/V$}. The overall factor of $i$ before the quadratic interaction term derives
from the phase correlation of counter-propagating modes in the ring. This factor may be absorbed
via a redefinition of operators, $\hat a_k\rightarrow \exp(-i\pi/4) \hat a_k$. This leads (via a diagonalization) to a set of eigenvalues corresponding to
an effective Hamiltonian given by $\epsilon^r_k=\sqrt{(\epsilon_k- J+gn_0- gn_r) ^2-g^2n_r^2}$. Here we consider the limit $n_r=n_0$, where nearly all particles are transferred to the
contribution from $n_r$.

The key point is that the main consequence of these ring-ring scattering events is
to scatter the particles in the ring back to momenta around $k\simeq 0$.
This is the counterpart of the cosmological
``back-reaction",
effect which stabilizes
a co-existing BEC (corresponding to simultaneous
ring and $k=0$ condensates).
Ultimately in the ``re-scattering" phase, collisions
between the two lead to a de-stablization.
In summary, on a time scale $t\sim t_r$ which corresponds to when the ring BEC has been formed, there will
be a counter-flow from the ring BEC back to modes at the band minima, until a balance of flow is reached between two BECs.

\subsection{Re-scattering: formation of the important cloud state}

In this section, we establish a mechanism to destabilize the co-existing BEC's, leading
to the emergence of a cloud state which represents a proliferation of bosonic pairs.

As in the previous sections, one can write down the scattering terms between the co-existing BECs,
\begin{eqnarray}
\hat{V}_{0,r}&\equiv&\frac{g}{V}   \left( \sum_{\mathbf{k}\in \Lambda,\mathbf{q}\notin \Lambda}  \hat a_{\mathbf{q}}^\dagger \hat a^\dagger_{\mathbf{k}-\mathbf{q}}\hat a_{\mathbf{0}} \hat a_{\mathbf{k}} +2\sum_{\mathbf{k}\in \Lambda} \hat a^\dagger_{\mathbf{0}} \hat a^\dagger_{\mathbf{k}}  \hat a_{\mathbf{0}} \hat a_{\mathbf{k}} + 2\sum_{\mathbf{k}\in \Lambda,\mathbf{q}\notin \Lambda} \hat a_{\mathbf{0}}^\dagger \hat a^\dagger_{\mathbf{q}}\hat a_{\mathbf{q}-\mathbf{k}} \hat a_{\mathbf{k}}+ 2\sum_{\mathbf{k}\in \Lambda,\mathbf{q}\notin \Lambda}  \hat a_{\mathbf{k}}^\dagger \hat a^\dagger_{\mathbf{q}-\mathbf{k}}\hat a_{\mathbf{q}} \hat a_{\mathbf{0}}  \right) \nonumber\\
\end{eqnarray}

\begin{figure*}[h]
            \centering
            \subfloat[][Imaginary part]{\includegraphics[scale=0.5]{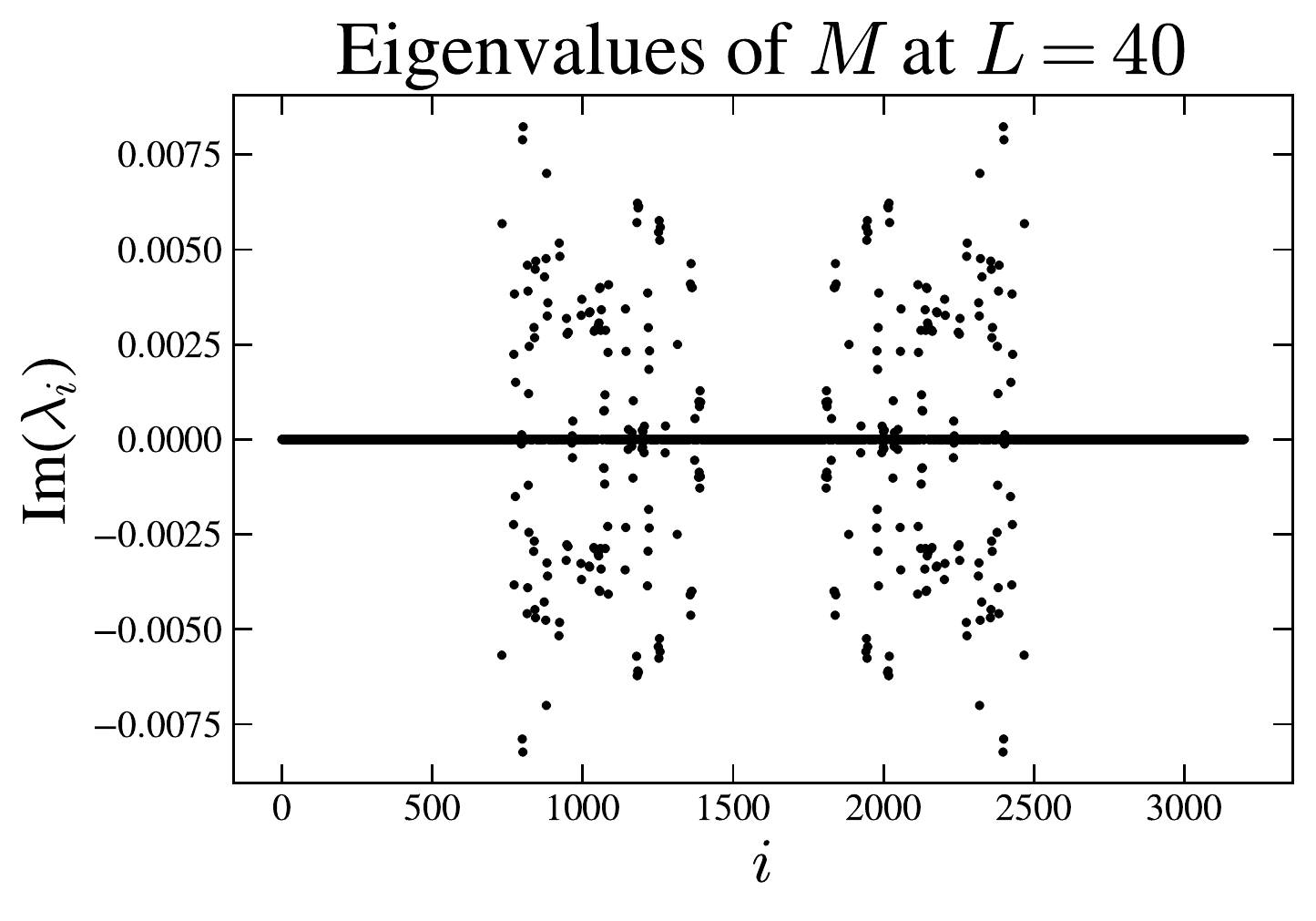}}
            \subfloat[][Real part]{\includegraphics[scale=0.5]{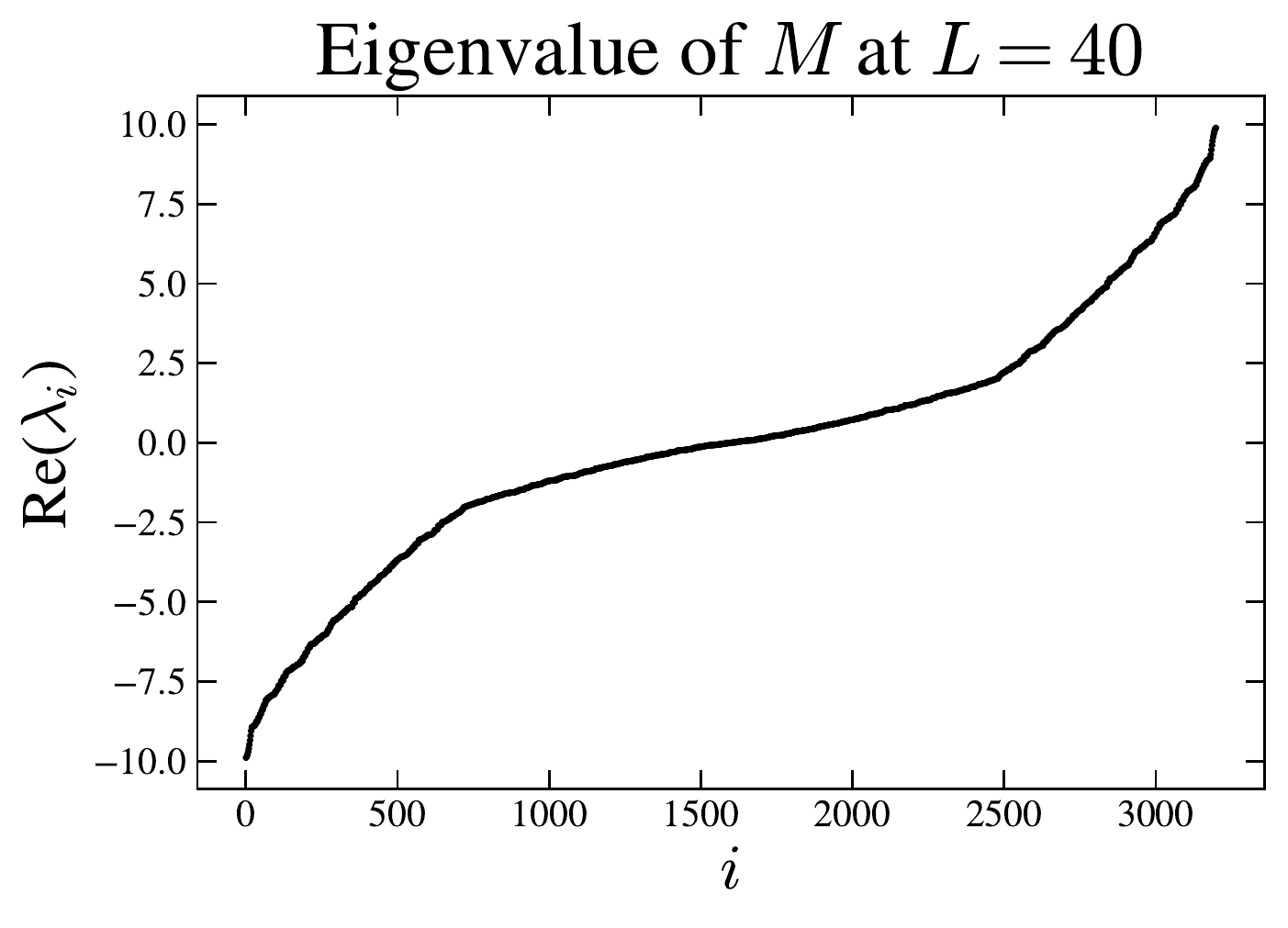}}
            \caption{
            Eigenvalue of the effective Hamiltonian in Eq.~\ref{11}. System size is set to be 40. The figure (a) plots the imaginary part of eigenvalues. One can observe there are a large number of eigenvalues with a small but finite imaginary part. The figure (b) plots the real part of the eigenvalues. Combining these two figures, one may 
deduce that there are a large number of complex eigenvalues.}
            \label{eigen}
        \end{figure*}

After a mean field approximation we
arrive at an effective Hamiltonian describing scattering events among co-existing BECs and other modes which leads to Eq.~(4) in the main text:
\begin{eqnarray}
\label{11}
\hat{H}^{0,r}_{\text{eff}}&\equiv&\sum_{ \mathbf{k}\notin \Lambda } (\epsilon_\mathbf{k}- J+gn_0/2-  g\bar{n}_r/2
+g\bar{n}_0/2) \hat a^\dagger_{ \mathbf{k}} \hat a_{ \mathbf{k} }  \\
&+&
2g\sqrt{\frac{\bar{n}_0 \bar{n}_r}{L_r}} \sum_{\mathbf{q}\in \Lambda,\mathbf{k}\notin \Lambda } ( \hat a_{\mathbf{k}}^\dagger \hat a^\dagger_{\mathbf{q}-\mathbf{k}}+h.c.  )+2g\sqrt{\frac{\bar{n}_0 \bar{n}_r}{L_r}}\sum_{\mathbf{q}\in \Lambda,\mathbf{k}\notin \Lambda } (  \hat a_{\mathbf{k}}^\dagger \hat a_{\mathbf{k}-\mathbf{q}}+h.c. ) \nonumber
\end{eqnarray}
Here $\bar n_0=N_0/V$ is the density of $k=0$ particles. (Note that the
random phase from $\hat a_\mathbf{q}$ is unimportant as it can be absorbed into redefined $\hat a_\mathbf{k}$ operators.)
Importantly the scattering events here are responsible for states appearing in
a new momentum regime.
The Hamiltonian is intrinsically non-local in momentum space and the exact treatment involves a
diagonalization of a matrix $M$ which includes multiple degrees of freedom.
this leads to an equation of motion for $a_k$
\begin{eqnarray}\label{eq:linear}
 \frac{d}{dt} \hat a_\mathbf{k}(t)&=& {\frac{-i}{\hbar}}\Bigg(
 \epsilon''_\mathbf{k} \hat a_\mathbf{k}+ 2g n_{\text{geom}}
 \sum_{\mathbf{q}\in \Lambda} \frac{1}{\sqrt{L_r}}  \left(\hat  a_{\mathbf{q}+\mathbf{k}}(t)+\hat a^\dagger_{\mathbf{q}-\mathbf{k}}(t) \right) \Bigg)
 \end{eqnarray}
 Here $\epsilon''_\mathbf{k} =\epsilon_\mathbf{k}- J+gn_0/2-  g\bar{n}_r/2
+g\bar{n}_0/2$ and $n_{\text{geom}}=\sqrt{\bar{n}_0 \bar{n}_r}$.
The closed set of equations for all $k$ modes can be compactly written as
 \begin{eqnarray}
\frac{d}{dt} \Psi(t)= {\frac{-i}{\hbar}} M \Psi(t)  \iff \Psi(t)=\exp \{ {-i M t/\hbar} \} \Psi(0).
\end{eqnarray}
where the vector $\Psi$ is defined by
\begin{eqnarray}
\Psi(t)= \begin{pmatrix}
\hat a_{\mathbf{k}_{1,1}}(t) &
 \hat a^\dagger_{-\mathbf{k}_{1,1}}(t) &
 ... &
 ... &
\hat a_{\mathbf{k}_{L,L}}(t) &
\hat a^\dagger_{-\mathbf{k}_{L,L}}(t)
\end{pmatrix}^T, \quad \mathbf{k}_{i,j}=\frac{2\pi}{L} (i,j),
 \end{eqnarray}
and $L$ is the linear size of system.

Numerically diagonalizing Eq.~\ref{eq:linear}, we find that the spectrum consists of a large number of
of {\it complex} eigenvalues.
(See Fig~\ref{eigen}). This can be simply physically interpreted.
In particular, since each non-local scattering term has a small coefficient, one can deal with each pair of $(\mathbf{q}, \mathbf{k}-\mathbf{q})$ individually. One may write down the
equation of motion for a single pair of momenta (as in Eq.~(5) of main text):
\begin{eqnarray}
\label{Eom}
\frac{d}{dt} \begin{bmatrix}
\hat a_\mathbf{k}(t) \\
 \hat a^\dagger_{\mathbf{q}-\mathbf{k}}(t)
\end{bmatrix}=   {\frac{-i}{\hbar}}\begin{bmatrix}
\epsilon''_\mathbf{k} & U'     \\
- U'  & -\epsilon''_{\mathbf{q}-\mathbf{k}}
\end{bmatrix}\begin{bmatrix}
\hat a_\mathbf{k}(t) \\
 \hat a^\dagger_{\mathbf{q}-\mathbf{k}}(t)
\end{bmatrix}.
\end{eqnarray}
Here $\epsilon''_\mathbf{k}=\epsilon_\mathbf{k}- J+g{n}_0/2- g\bar{n}_r/2+ g\bar{n}_0/2$ and $ {U'}\equiv 2g\sqrt{\frac{\bar{n}_0 \bar{n}_r}{L_r}}$.
The eigenvalue of the $2\times 2$ matrix above is given by $
\lambda=\frac{\epsilon''_\mathbf{k}-\epsilon''_{\mathbf{k}-\mathbf{q}}}{2}\pm \sqrt{
\left( \frac{\epsilon''_\mathbf{k}+\epsilon''_{\mathbf{q}-\mathbf{k}}}{2} \right)^2
- {U'}^2
}.
$
This complex solution reflects exponential growth. Hence, the fastest growing pair is such that
$\epsilon''_\mathbf{k}=-\epsilon''_{\mathbf{k}-\mathbf{q}} \iff  \epsilon_\mathbf{k}+\epsilon_{\mathbf{k}-\mathbf{q}} =2(J-g\bar{n}_0)
$, assuming $n_0=\bar{n}_0+\bar{n}_r$.

This analysis can be used to determine the shape of the cloud. Consider the following two equations,
 \begin{eqnarray}
&&J \cos  {(k_yd)}+ \frac{\hbar^2}{2m} {k^2_x}-(J-g \bar{n}_0 )=(J-g \bar{n}_0 )-J {\cos\left[ (k_y-q_y)d\right]}-\frac{\hbar^2}{2m}  {(k_x-q_x)^2} \nonumber \\
&& \text{ with } \quad J\cos {( q_yd)} +  \frac{\hbar^2}{2m} { q^2_x}=J-g {n_0} .
 \end{eqnarray}
 The first equation introduces a constraint on the kinetic energy and second equation
determines the ring shape. To further simplify these equations,
one may cancel $q_y$ in the first and second equations and find a single equation below,
\begin{eqnarray}
0&=&J\left(2- {\frac{\hbar^2q_x^2}{2mJ}-\frac{g n_0}{J}}\right) \cos  {(k_yd)}-2(J-g \bar{n}_0) \\
&\pm& J\sin {( k_yd)} \sqrt{1- {\left(1-\frac{\hbar^2 q_x^2}{2mJ}-\frac{g n_0}{J}\right)^2}}+  {\frac{\hbar^2}{2m}\left(2k_x^2-2k_xq_x+q_x^2\right)} \equiv F_{ {q_x}}(\mathbf{k}) .\nonumber
\end{eqnarray}
It follows that the cloud shape $C$ is determined by zeros of the functions $F_{ {q_x}}(\mathbf{k})$, where
$q_x$ is a free parameter.
Fig.~3b in the main text is obtained by solving for zeros of these functions numerically.

It is important to stress that after the reheating stage is completed all remnants of
`global' coherence decay away at which point the cloud state is fully formed. For the isolated quantum system we study, quantum information cannot be lost, but it can become spread and scrambled in a more complicated manner.
Note that the cloud state is a non-thermal state. Nevertheless we discuss next how it enables the system to thermalize.


\section{Thermalization: Boltzmann dynamics}

During this last stage of equilibration, the cloud state reorganizes and the collection
of non thermal bosons contained in the cloud begin to occupy
the band minimum. This process is associated with Boltzmann dynamics, as it involves no `global' phase coherence.
Since we consider weak interactions, one may use leading order perturbative theory to characterize the processes.
The dynamical equation of each $k$-mode is given by
\begin{eqnarray}
\frac{\partial n_\mathbf{k}}{\partial t}= \frac{2g^2}{\hbar} \int \frac{d^2 \mathbf{k}' d^2\mathbf{q}}{(2\pi)^4} \Gamma(\mathbf{k},\mathbf{k}';\mathbf{k}'+\mathbf{q},\mathbf{k}-\mathbf{q}) 2\pi \delta(\epsilon_\mathbf{k}+\epsilon_{\mathbf{k}'}-\epsilon_{\mathbf{k}+\mathbf{q}}-\epsilon_{\mathbf{k}'-\mathbf{q}}).
\end{eqnarray}
Here the vertex function $\Gamma$ determines how the particle number is changed for scattering
events involving the four momenta: $\mathbf{k},\mathbf{k}',\mathbf{k}+\mathbf{q},\mathbf{k}-\mathbf{q}'$:
\begin{eqnarray}\label{eq:bolzmann}
\Gamma(\mathbf{k},\mathbf{k}';\mathbf{k}+\mathbf{q},\mathbf{k}-\mathbf{q}')=n_{\mathbf{k}-\mathbf{q}} n_{\mathbf{k}'+\mathbf{q}} (1+n_\mathbf{k}) (1+n_{\mathbf{k}'})- (n_{\mathbf{k}-\mathbf{q}}+1) (n_{\mathbf{k}'+\mathbf{q}}+1)  n_\mathbf{k}  n_{\mathbf{k}'}. 
\end{eqnarray}
The scattering event $\mathbf{k},\mathbf{k}' \rightarrow \mathbf{k}-\mathbf{q}, \mathbf{k}'+\mathbf{q} $ decreases $n_\mathbf{k}$ by unity while the reverse process
similarly increases $n_\mathbf{k}$. Note that $\mathbf{k},\mathbf{k}' \rightarrow \mathbf{k}-\mathbf{q}, \mathbf{k}'+\mathbf{q} $ and $\mathbf{k}-\mathbf{q},\mathbf{k}'+\mathbf{q} \rightarrow \mathbf{k}, \mathbf{k}' $ have different scattering amplitudes. Their difference leads to a cubic power dependence on the particle number. Thus,
\begin{eqnarray}
\frac{\partial n_\mathbf{k}}{\partial t}=I_\mathbf{k}(n).
\end{eqnarray}
This equation reflects three conserved quantities: particle number, momentum, and energy.

Additionally, this Boltzmann dynamics reflects the H-theorem which is associated with
a monotonic increase in the entropy of the Bose gas discussed in Fig.~4 of the
main text and given by
\begin{eqnarray}
S_B=\int d^2\mathbf{k}\left[  (n_\mathbf{k}+1) \ln (n_\mathbf{k}+1)-     n_\mathbf{k}  \ln  n_\mathbf{k}    \right].
\end{eqnarray}

\subsection{Characterization of cloud state.}

To understand the equilibrating Boltzmann dynamics, it is important to provide a
description of the initial state, the cloud state $|\Psi_\Omega \rangle$. This
is the intermediate non-thermal state which appears after the quenching of
the condensate but before thermal equilibrium. It satisfies the following:
\begin{itemize}
    \item It is contained in a simply-connected space $\Omega $ with well defined
boundaries in momentum space. $\Omega$ is, thus, a sub-space of the two dimensional Brillouin zone (BZ).
Importantly the smaller the size of $\Omega$ the lower the kinetic energy of the system and
hence the lower its effective temperature.

\item For those momentum states which are present there is an effectively
large occupation number $n_k(\Omega)=\langle \Psi_\Omega| \hat n_k | \Psi_\Omega\rangle$.
Indeed in our numerical simulations we find
$n_k(\Omega)\gg 1$.
\item Within these occupied states
there is no `global' phase coherence
    \begin{eqnarray}
     \langle \Psi_C | \hat a_k \hat a_{-k} | \Psi_C \rangle \ll 1.
    \end{eqnarray}
The absence of phase coherence, means that the cloud state can be treated within a Boltzmann equation
approximation.
\end{itemize}


\subsection{The early stage of Boltzmann dynamics}

The dynamics in this early stage strongly depends on these properties of the cloud state.
One may classify the 2-body collision events according to how many particles inside
and outside the cloud are involved. If there are 3 or 4 inside-the-cloud modes involved
this will induce strong particle number flow, whereas if there are only
2 such modes the flow is weaker, and with only one or zero such modes, this contribution
in the early dynamics can be ignored.

We focus, for simplicity, on a situation where the particle number distribution in the cloud is uniform.
We have tested the appropriateness of this assumption numerically and
with this assumption, one can
greatly simplify the expression for the collision integral. To this end, we define four functions below,
\begin{eqnarray}
F_1(\mathbf{k})=\int_{\mathbf{k}_2 \notin C} \frac{d^2 \mathbf{k}_2 }{(2\pi)^2}  \int_{\mathbf{k}_3 \in C} \frac{d^2 \mathbf{k}_3 }{(2\pi)^2}  \int_{\mathbf{k}_4 \in C} \frac{d^2 \mathbf{k}_4 }{(2\pi)^2}   2\pi \delta(\epsilon_{\mathbf{k}}+\epsilon_{\mathbf{k}_2}-\epsilon_{\mathbf{k}_3}-\epsilon_{\mathbf{k}_4})   {\left(2\pi\right)^2}  \delta(\mathbf{k}+\mathbf{k}_2-\mathbf{k}_3-\mathbf{k}4), \nonumber \\
F_2(\mathbf{k})=\int_{\mathbf{k}_2 \in C} \frac{d^2 \mathbf{k}_2 }{(2\pi)^2}  \int_{\mathbf{k}_3 \notin C} \frac{d^2 \mathbf{k}_3 }{(2\pi)^2}  \int_{\mathbf{k}_4 \in C} \frac{d^2 \mathbf{k}_4 }{(2\pi)^2}   2\pi \delta(\epsilon_{\mathbf{k}}+\epsilon_{\mathbf{k}_2}-\epsilon_{\mathbf{k}_3}-\epsilon_{\mathbf{k}_4})   {\left(2\pi\right)^2}  \delta(\mathbf{k}+\mathbf{k}_2-\mathbf{k}_3-\mathbf{k}4), \nonumber \\
G(\mathbf{k})=\int_{\mathbf{k}_2 \in C} \frac{d^2 \mathbf{k}_2 }{(2\pi)^2}  \int_{\mathbf{k}_3 \notin C} \frac{d^2 \mathbf{k}_3 }{(2\pi)^2}  \int_{\mathbf{k}_4 \notin C} \frac{d^2 \mathbf{k}_4 }{(2\pi)^2}   2\pi \delta(\epsilon_{\mathbf{k}}+\epsilon_{\mathbf{k}_2}-\epsilon_{\mathbf{k}_3}-\epsilon_{\mathbf{k}_4})   {\left(2\pi\right)^2}  \delta(\mathbf{k}+\mathbf{k}_2-\mathbf{k}_3-\mathbf{k}_4). \nonumber\\
J(\mathbf{k})=\int_{\mathbf{k}_2 \in C} \frac{d^2 \mathbf{k}_2 }{(2\pi)^2}  \int_{\mathbf{k}_3 \in C} \frac{d^2 \mathbf{k}_3 }{(2\pi)^2}  \int_{\mathbf{k}_4 \in C} \frac{d^2 \mathbf{k}_4 }{(2\pi)^2}   2\pi \delta(\epsilon_{\mathbf{k}}+\epsilon_{\mathbf{k}_2}-\epsilon_{\mathbf{k}_3}-\epsilon_{\mathbf{k}_4})   {\left(2\pi\right)^2}  \delta(\mathbf{k}+\mathbf{k}_2-\mathbf{k}_3-\mathbf{k}_4), \nonumber
\end{eqnarray}

We consider the two separate cases for
$\mathbf{k}\in \Omega$ and
$\mathbf{k}\notin \Omega$. For the former
the collision integral can be reduced to be,
\begin{eqnarray}
I_{\bf{k}}(n)\simeq  \frac{2g^2}{\hbar}  \left[F_1(\mathbf{k}) n^3_\Omega -2 F_2(\mathbf{k})n^3_\Omega- G(\mathbf{k}) n_\Omega^2   \right],
\end{eqnarray}
where only the first two terms dominate.
Moreover for
$\mathbf{k}\notin \Omega$
the collision integral becomes,
 \begin{equation}
I_\mathbf{k}(n)\simeq  \frac{2g^2}{\hbar}  \left[J(\mathbf{k}) n^3_\Omega   + F_1(\mathbf{k})  {n^2_\Omega}  \right].
\end{equation}
Here numerically one can show that both $J(\mathbf{k})$ and $F_1(\mathbf{k})$ nearly vanish except in a small region near the boundary of the cloud.



\section{Results based on the Gross-Pitaevskii equation}

To connect more directly to our numerical GP simulations it is useful to consider the semi-classical approximation of the quantum theory in Eq.~\ref{upper} based on the GP equation:
\begin{eqnarray}
\label{1}
\left[J \cos(i {\partial_{y}d})- {\frac{\hbar^2\partial^2_{x}}{2m}} + g|\psi(\mathbf{r},t)|^2\right] \psi(\mathbf{r},t)=i  {\hbar} \frac{\partial}{\partial t} \psi(\mathbf{r},t).
\end{eqnarray}

The evolutionary dynamics found in our numerical GP simulations is seen to be consistent with
the analytics presented in the previous sections.
Also consistent is the close tie to observations from cosmological models. We can demonstrate this through
a linearization of the GP equation appropriate to the preheating stage.


In this early stage dynamics (associated with the preheating phase in cosmological models), the physics is governed by a parametric
resonance.
The initial quenched condensate with a finite kinetic energy can be modeled by the uniform and intrinsically oscillating
scalar field,
$
\psi_0(\mathbf{r},t)=\sqrt{n_0} e^{-i \omega t }.
$
Here $\omega$ is the intrinsic oscillating frequency and its value is determined by solving the corresponding
GP equation, as we show below. We assume a
wavefunction of the form, $\psi(\mathbf{r},t)= \psi_0({\bf{r}},t)+\delta \psi(\mathbf{r},t)$.
To linearize the theory, we treat the self-interaction term perturbatively,
\begin{eqnarray}
|\psi(\mathbf{r},t)|^2=\left | \sqrt{\frac{N}{V}} e^{-i \omega t } +\delta \psi \right|^2 \simeq n_0+ \sqrt{\frac{N}{V}} e^{-i\omega t }  \delta \psi^* + h.c. + O(|\delta \psi|^2)
\end{eqnarray}
Now we write down the equation for $\delta \psi$. Using the expansion variable $\delta \psi$,
it follows that
\begin{equation}
\label{S17}
\left[J \cos(i {\partial_y d})- {\frac{\hbar^2\partial^2_x}{2m}} + gn_0+ g\sqrt{\frac{N}{V}}e^{-i \omega t } \delta \psi^*+ g \sqrt{\frac{N}{V}}e^{i \omega t } \delta \psi \right] \left(
\sqrt{\frac{N}{V}} e^{-i \omega t }+ \delta \psi
\right)= i {\hbar}\partial_t \delta \psi+ {\hbar}\omega \sqrt{\frac{ {N}}{V}} e^{-i \omega t }. \nonumber\\
\end{equation}
Since $\psi_0$ satisfies the GP equation, it follows that $${\hbar}\omega=J+gn_0, $$ indicating that $\omega$ is intrinsically determined by the model parameters.
Keeping only the linear terms in $\delta \psi, \delta \psi^*$, one finds
\begin{equation}
\left[J \cos(i {\partial_y d})- {\frac{\hbar^2\partial^2_x}{2m}} + 2gn_0   \right]\delta \psi+ g n_0 e^{-2i\omega t} \delta \psi^* =i {\hbar} \partial_t  \delta \psi.
\end{equation}

One may write the equation in the simple form,
\begin{equation}
E_K(-i\boldsymbol{\nabla}) \delta \psi+ g(t) \delta \psi^* =i {\hbar} \partial_t  \delta \psi.
\end{equation}
where $E_K(-i\boldsymbol{\nabla})=J \cos(i {\partial_y d})- {\frac{\hbar^2\partial^2_x}{2m}} + 2gn_0$ is the single-particle energy with the Hartree-Fock shift and $g(t)=gn_0 \exp(-2i\omega t)$ is the effective time-dependent interaction. The time-dependence originates from the fact that the condensate is intrinsically oscillating. Later we will see how the oscillating $g(t)$ leads to the parametric instability. Note that in
Ref.~\onlinecite{pra21Chatrchyan}
the modulation of the coupling (to be time-dependent) is imposed externally and the oscillating frequency serves as a free parameter of their model. As a contrast, the frequency $\omega$ is {\it intrinsic} in our model and has a clear physical origin, the oscillating frequency of the condensate.

Expanding $\delta \psi=
\delta \psi(\mathbf{r},t)=\sum_{\mathbf{k},\nu} e^{i\mathbf{k}\cdot\mathbf{r}-i\nu t} \psi_\mathbf{k} (\nu)$, 
Eq.~\ref{S17} then becomes
\begin{equation}
\left[J \cos( {k_yd}) +  {\frac{\hbar^2k_x^2}{2m}}+ 2gn_0   \right] \psi_\mathbf{k} (\nu)+ g n_0    \psi^*_{-\mathbf{k}} ( -\nu +2\omega )=  { \hbar}\nu  \psi_\mathbf{k}(\nu).
\end{equation}
With a change of variable  $\nu= \nu'+\omega$ we arrive at a matrix equation of the form
\begin{equation}
\left\{\left[J \cos( {k_yd}) + {\frac{\hbar^2k_x^2}{2m}}+ 2gn_0-\omega \right]  \hat\sigma_z+gn_0  \hat\sigma_z \hat\sigma_x \right\}
\begin{bmatrix}
\psi_\mathbf{k}(\nu'+\omega)\\
\psi^*_{-\mathbf{k}}(\omega-\nu')
\end{bmatrix}
=\nu'    \begin{bmatrix}
\psi_\mathbf{k}(\nu'+\omega)\\
\psi^*_{-\mathbf{k}}(\omega-\nu')
\end{bmatrix}  .
\end{equation}
Here  
$\hat\sigma_{x,y,z}$ is the Pauli matrix. Solving this equation, one finds
\begin{equation}
\nu {'}(\mathbf{k})= {\pm}\sqrt{ \left[J \cos( {k_yd}) + {\frac{\hbar^2k_x^2}{2m}}+ 2gn_0-\omega  \right]^2 -(g n_0)^2   }.
\end{equation}
When $\nu'$ assumes imaginary values, this corresponds to the parametric resonance
condition in cosmological models. When $\omega$ is taken to be $(J+gn_0) {/\hbar}$,
this is consistent with
Eq.~\ref{S6}.

In this way, we have shown the existence of a parametric instability in an isolated quantum system when the condensate has a finite intrinsic oscillating frequency.  The physical process behind the parametric resonance is that the energy of the condensate transfers to the non-condensate particles in an exponentially amplified way.

\subsection{GP derivation of the cloud thermalization stage: Comparison with the quantum Boltzmann equation}

It is instructive to use the machinery of the GP equation to characterize how equilibration
occurs. We will begin with a stage in which
there is no `global' phase coherence
and write 
$\psi(\mathbf{r},t)=\sum_{\bf{k}}\psi_\mathbf{k}(t)e^{i\mathbf{k}\cdot\mathbf{r}}$, the corresponding equation of motion in terms of the complex scalar field $\psi_{\mathbf{k}}(t)$ is then
\begin{equation}
i\hbar\frac{\partial \psi_\mathbf{k}(t)}{\partial t}=\epsilon_\mathbf{k}+g\sum_{\mathbf{k_1},\mathbf{k_3}}\psi_\mathbf{k_1}^*\psi_\mathbf{k_3}\psi_\mathbf{k_4}
\end{equation}
where $\mathbf{k}_4=\mathbf{k}+\mathbf{k_1}-\mathbf{k_3}.$ For simplicity, we put $\psi_\mathbf{k}(t)=\tilde{\psi}_\mathbf{k}(t)e^{-i\epsilon_\mathbf{k}t/\hbar}$. We then arrive at
\begin{equation}
\begin{aligned}
i\hbar\frac{\partial \tilde{\psi}_\mathbf{k}(t)}{\partial t}=g&\sum e^{i(\epsilon_i-\epsilon_f)t/\hbar}\tilde{\psi}_\mathbf{k_1}^*\tilde{\psi}_\mathbf{k_3}\tilde{\psi}_\mathbf{k_4}\\
\frac{\partial\left( \tilde{\psi}_\mathbf{k}^*\tilde{\psi}_\mathbf{k}\right)}{\partial t}=&\frac{-i}{\hbar}\tilde{\psi}_\mathbf{k}^*g\sum e^{i(\epsilon_i-\epsilon_f)t/\hbar}\tilde{\psi}_\mathbf{k_1}^*\tilde{\psi}_\mathbf{k_3}\tilde{\psi}_\mathbf{k_4}+h.c.
\end{aligned}
\end{equation}
where $\epsilon_i=\epsilon_\mathbf{k_3}+\epsilon_\mathbf{k_4},\,\epsilon_f=\epsilon_\mathbf{k}+\epsilon_\mathbf{k_1},\,n_\mathbf{k}=
\psi_\mathbf{k}^*\psi_\mathbf{k}=\tilde\psi_\mathbf{k}^*\tilde\psi_\mathbf{k}.$ We now assume that
in addition to the absence of `global' phase coherence, the interaction energy is quite small. As a result, the time change of $\tilde\psi_\mathbf{k}(t)$ can be approximated to the first order in $g$, and the change rate is much smaller than $\epsilon_i/\hbar,\epsilon_f/\hbar$. We then obtain
\begin{equation}
\begin{aligned}
\tilde\psi_\mathbf{k}(t)\approx&\tilde\psi_\mathbf{k}(0)-i\pi g\sum\delta(\epsilon_i-\epsilon_f)\tilde{\psi}_\mathbf{k_1}^*\tilde{\psi}_\mathbf{k_3}\tilde{\psi}_\mathbf{k_4}\\
\frac{\partial n_\mathbf{k}}{\partial t}=&\frac{-ig}{\hbar}\left[\tilde\psi_\mathbf{k}^*(0)+i\pi g \sum'\delta(\epsilon_i'-\epsilon_f')\tilde{\psi}_\mathbf{k_1'}\tilde{\psi}_\mathbf{k_3'}^*\tilde{\psi}_\mathbf{k_4'}^*\right]\sum e^{i(\epsilon_i-\epsilon_f)t/\hbar}\times\left[\tilde\psi_\mathbf{k_1}^*(0)+i\pi g \sum'\delta(\epsilon_i'-\epsilon_f')\tilde{\psi}_\mathbf{k'}\tilde{\psi}_\mathbf{k_3'}^*\tilde{\psi}_\mathbf{k_4'}^*\right]\\
&\times\left[\tilde\psi_\mathbf{k_3}(0)-i\pi g \sum'\delta(\epsilon_i'-\epsilon_f')\tilde{\psi}_\mathbf{k_4'}^*\tilde{\psi}_\mathbf{k'}\tilde{\psi}_\mathbf{k_1'}\right]\times\left[\tilde\psi_\mathbf{k_4}(0)-i\pi g \sum'\delta(\epsilon_i'-\epsilon_f')\tilde{\psi}_\mathbf{k_3'}^*\tilde{\psi}_\mathbf{k'}\tilde{\psi}_\mathbf{k_1'}\right]+h.c.
\end{aligned}
\end{equation}
where again both momentum and energy conservation are maintained. Now because of the lack of phase coherence, different $\tilde\psi$ amplitudes have uncorrelated random phases. Only terms with both $\psi$ and $\psi^*$ of the same momentum will survive in the equation. Therefore, we find
\begin{equation}
\begin{aligned}
\frac{\partial n_\mathbf{k}}{\partial t}=&\frac{-ig}{\hbar}\left[\sum 2|\tilde\psi_\mathbf{k}|^2|\tilde\psi_\mathbf{k_1}|^2\right.+2\pi ig \sum |\tilde\psi_\mathbf{k_1}|^2|\tilde\psi_\mathbf{k_3}|^2|\tilde\psi_\mathbf{k_4}|^2\delta(\epsilon_i-\epsilon_f)+2\pi ig \sum |\tilde\psi_\mathbf{k}|^2|\tilde\psi_\mathbf{k_3}|^2|\tilde\psi_\mathbf{k_4}|^2\delta(\epsilon_i-\epsilon_f)\\
&-2\pi ig \sum |\tilde\psi_\mathbf{k}|^2|\tilde\psi_\mathbf{k_1}|^2|\tilde\psi_\mathbf{k_4}|^2\delta(\epsilon_i-\epsilon_f)\left.-2\pi ig \sum |\tilde\psi_\mathbf{k}|^2|\tilde\psi_\mathbf{k_1}|^2|\tilde\psi_\mathbf{k_3}|^2\delta(\epsilon_i-\epsilon_f)\right]+h.c.\\
&=\frac{4\pi g^2}{\hbar}\left(n_\mathbf{k_1}n_\mathbf{k_3}n_\mathbf{k_4}+n_\mathbf{k}n_\mathbf{k_3}n_\mathbf{k_4}-n_\mathbf{k}n_\mathbf{k_1}n_\mathbf{k_3}-n_\mathbf{k}n_\mathbf{k_1}n_\mathbf{k_4}\right),
\end{aligned}
\end{equation}
where higher orders in $g$ are discarded.
Comparing this result with Eq.~\eqref{eq:bolzmann}, we find that the term $n_\mathbf{k_3}n_\mathbf{k_4}-n_\mathbf{k}n_\mathbf{k_1}$ is missing in this approximate GP approach.
Thus, the GP approach works best in the early stages of thermalization where
the leading contribution is $\propto n^3$. We have chosen a large particle density for our simulations to ensure this dominance of $n^3$ in the early stage.
We infer that there may be some discrepancies at very late times.



\section{numerical parameters in the Gross-Pitaevskii simulations}

The parameters we used for the simulations are: $V_1=12 E_R, V_2=2E_R$, $gn_0=0.015E_R$, $h/T=4.75 E_R$ where $E_R=\hbar^2(\pi/d)^2/2m$ is the recoil energy unit. $d$ is the period of the lattice, which is discretized into 64 grids in the $y$-direction with a total length of $256d$. For the $x$-direction which is free and thus insensitive to grid resolution, we use $64d$ for the total length to speed up the simulations. Periodic boundary conditions are imposed for all directions. We run the dynamics of the system 
for a sufficiently long time until we find the results are stable. 
The longest running time is 6000T.

Additionally, random noise terms have been added to the system to simulate quantum fluctuations. 
Our simulations involve
 Graphics Processing Unit-based
quasispectral, split-step method to solve the GP equation based on fast Fourier
transforms.
More details can be found in ``Pseudo-spectral solution of
nonlinear Schroedinger equations" in the Journal of Computational Physics 87,
pages 108-125 (1990).

\bibliography{document1}